\documentclass[preprint2]{aastex}
\usepackage{graphicx}

\slugcomment{Submitted to Astronomical Journal}

\shorttitle{[Ar III] Knots in the Crab Nebula} 
\shortauthors{Schaller \& Fesen}
\begin{document}

\title{The Nature of [Ar III] Bright Knots in the Crab Nebula}

\author{Emily L. Schaller \& Robert A. Fesen}
\affil{6127 Wilder Laboratory, Department of Physics \& Astronomy \\
       Dartmouth College, Hanover, NH 03755} 

\begin{abstract}

The kinematic and morphological properties of a string of [Ar~III] bright knots 
in the Crab Nebula are examined using 1994 -- 1999 
{\it HST} WFPC-2 images of the remnant. 
We find that five southern [Ar~III] bright knots exhibit 
ordinary radial motions away from the nebula's center of expansion 
with magnitudes consistent with their projected radial displacements. 
This result does not support the suggestion by \citet{MacAlpine94} 
that these knots might be moving rapidly away from the Crab pulsar 
due to a collimated wind. 
The {\it HST} images also do not show that the [Ar~III] knots  
have unusual morphologies relative to other features in the remnant. 
Our proper motion results, when combined with radial velocity estimates,
suggest these knots have relatively low space velocities implying relatively 
interior remnant locations thus placing them closer to the ionizing radiation from the
Crab's synchrotron nebula. This might lead to higher knot gas temperatures
thereby explaining the knots' unusual line emission strengths 
as \citet{MacAlpine94} suspected.

\end{abstract}

\section{Introduction}

For over half a century, the Crab Nebula has played a key role as a laboratory for many 
seminal discoveries regarding supernovae and their remnants \citep{Trimble85}.
It is also the brightest and best studied example of ``plerionic'' remnants which are 
powered by a compact central object (see reviews \citealt{KH85,DF85}).

Despite its many astronomical firsts, the detection by \citet{MacAlpine94} 
of chains of semi-stellar, optical knots showing unusually strong [Ar~III] line emission 
was both remarkable and puzzling. The discovery 
came about through a search looking for north-south, bi-polar 
axial phenomena directly associated with the Crab pulsar and  
related to the remnant's east-west band of strong helium emission filaments
\citep{UM87,MMMU89,Fesen92} and the N-S hourglass structure seen in  
polarized light of the Crab's synchrotron nebula \citep{Michel91}.

Using a Fabry-Perot imager centered at 5015.3 \AA, \citet{MacAlpine94} found
about a dozen, semi-stellar [O~III] $\lambda$5007 emission knots aligned in
arcs, perhaps helical \citep{MacAlpine92}, from the pulsar's position with 
seven to the north and four to the south.
The northern arc of knots appeared situated inside a corridor
through the remnant's filamentary structure, the eastern edge of
which seemed to merge with the western edge of the Crab's well-known
northern ``jet'' \citep{vdb70}.
\citet{MacAlpine94} suggested that this corridor containing
the northern knots marked the presence of a directed, north-south  
collimated wind from the pulsar's vicinity and possibly associated
with the pulsar's spin axis \citep{MacAlpine92}.

Subsequent optical spectra of several of these knots revealed 
remarkably strong [Ar III] $\lambda$7136 and 
[S~II] $\lambda\lambda6716,6731$ line emission.
The knots' strong [Ar~III] line emission was deemed 
more remarkable than that of the [S~II],
leading MacAlpine et al. to refer them as ``argo-knots''.
Their unusual [Ar~III] line strength was interpreted as being due 
to either N(Ar$^{+2}$)/N(H$^{+}$) ratios $5 - 10$ higher than in 
typical Crab filaments, unusually high knot electron temperatures, 
or some process(es) not normally important in nebula astrophysics. 

Perhaps most remarkable was the knots' apparent alignment 
with, and transverse velocities as high as 900 km s$^{-1}$ away from, the Crab pulsar. 
The knots' arrangement in north and south arcs in striking 
alignment with the Crab pulsar was highly suggestive of a causal link.
This is because the pulsar has a motion toward the northwest away from the 
Crab's expansion point of $\simeq$125 km s$^{-1}$
($0 \farcs 011 \pm 0 \farcs 001$ yr$^{-1}$,
PA = 290$^{\circ}$; \citealt{WM77}).
In addition to the location of the northern knots inside a filamentary
corridor associated with the northern jet feature, 
\citet{MacAlpine94} also found that the argo-knots' north-south alignment to be 
roughly perpendicular to an apparent east-west torus 
of high-helium filaments \citep{UM87}.

Positional measurements for the seven northern and four
southern [Ar~III] knots using ground-based images taken 
over a two year baseline (1991 -- 1993)
indicated proper motions of $\leq 0 \farcs 1$ yr$^{-1}$ or
$\leq$ 900 km s$^{-1}$ at the Crab's 1830 pc distance \citep{DF85}.
Although their proper motion results had relatively large associated 
uncertainties ($\pm0 \farcs 07$ yr$^{-1}$), \citet{MacAlpine94}
concluded that ``All measured north or south motions are 
directed away from the pulsar, as expected''.

This conclusion, when taken together with the knots'
remarkable alignment  with the pulsar's current position 
suggested a kinematic link between the pulsar and these so-called ``argo-knots''. 
However, if these knots were dynamically connected 
with a collimated wind off the pulsar, it would then mean that 
they comprise a small but previously 
unrecognized population of line emitting material 
within the remnant which does not participate in the 
remnant's radial expansion from the SN~1054 
explosion point \citep{Trimble68,Nugent98}. 

Non-radial motions in the Crab Nebula and in particular 
the presence of high-velocity, line-emitting gas moving 
rapidly away north and south from the pulsar 
is an unexpected discovery meriting further examination. 
In this paper, we present a proper motion and morphological 
study of a few of these knots using 1994-1999 {\it HST} WFPC-2 
images of the Crab Nebula.

\section{Observations}  

{\it HST} WFPC-2 images of the center of 
the Crab Nebula from 1994, 1995 and 1999 were
obtained from the data archive at the Space Telescope Science Institute (Table 1).
Figure 1 shows a 1995 {\it HST} WFPC2 image of the west-central region 
of the Crab Nebula taken with the [O~III] $\lambda$5007 filter (F502N). 
The strong [Ar III] emitting knots N1 -- N6 and S1 -- S4
described by \citealt{MacAlpine94} are marked. 
Other nebular knots having similar ``semi-stellar''
[O~III] morphologies are shown in the image and are labeled K1 -- K9. 

Because all 11 bright [Ar~III] knots exhibit prominent [O~III] $\lambda$5007 and
[S~II] $\lambda\lambda$6716,6731 line emissions and positive radial velocities
(+300 to +720 km s$^{-1}$; \citealt{MacAlpine94}),
WFPC-2 images taken using the [O~III] F502N and 
[S~II] F673N filters are well suited for
studying these knots. Both filter bandpass centers are slightly redshifted
(+360 km s$^{-1}$) and sufficiently wide ($\pm800$ km s$^{-1}$) to cover
the knots' observed radial velocity range. In addition, the [O~III] F502N image
line center at 5013 \AA \ is almost identical to the 5015 \AA \ Fabry-Perot bandpass
center used by \citet{MacAlpine94} to measure the [Ar~III] knots' kinematic
properties.

Pairs of WFPC-2 1994 and 1999 images taken in [O~III] and [S~II]
were each combined using the cosmic ray cleaning IRAF package ``crrej''.  
For the 1995 data, two pairs of images were available. 
These were also initially combined using ``crrej''
and then one pair was shifted and they were combined again using ``crrej''.  
The combined images for each year were then made into three 
mosaics representing 1994, 1995, and 1999, and rotated so that 
all were oriented with north up and east to the left.
  
Pixel image positions of four to six stars present on the 1994, 1995, and 1999 
images were measured on each of the [O~III] or [S~II] 
WFPC-2 images ($0 \farcs 0996$ pixel$^{-1}$) 
using the IRAF task ``imexamine''.  
Image positions for the four southern [Ar~III] bright
nebular knots identified by \citep{MacAlpine94} 
were measured both by eye and by using ``imexamine'',
with good agreement between the two methods. 
Since southern Knots S3 and S4 were not imaged on the 
1994 images, their proper motion values
were determined by comparing only the 1995 and 1999 images.
Also measured was a similar but somewhat fainter, neighboring feature to 
Knots S1 and S2 which we call ``S0'' due to its smaller, 
projected distance from the pulsar than Knot S1.
Pixel x and y offsets between the 1994/95 and 1999 images 
were then obtained and used to predict the 1999 pixel 
positions of the knots from the 1994 and 1995 images. 
Observed versus predicted knot positions 
divided by the time interval yielded $\alpha$ and $\delta$ (x and y) 
proper motion values.

Image pixel x and y differences for the reference stars 
were each averaged and the standard deviation 
of each was used as an estimate of the
minimum measurement errors. 
These were added in quadrature with error estimates
based on position differences 
between the [O~III] and [S~II] knots.
Because of the large and irregular morphology of Knot S3, 
its proper motion determination is significantly more uncertain.

\section{Results and Discussion}

The Crab pulsar currently lies some $12'' - 13''$ northwest of the
remnant's estimated expansion center \citep{Trimble68,Nugent98}.
To determine whether, as \citet{MacAlpine94} concluded, 
[Ar~III] bright knots move away from the 
the Crab pulsar ``as expected'' due to a presumed collimated wind off the pulsar
or instead move away the remnant's expansion center, 
we chose to examine the four southern [Ar~III]
bright knots they identified (S1--S4). We also included one  
additional likely [Ar~III] knot 
(S0) we identified from the WFPC-2 [O~III] images. 
This latter knot may be the one \citet{MacAlpine94} referred to as 
being simply ``to the south'' of the pulsar.

Due to locations south{\it east} of the pulsar but 
south{\it west} of the Crab's center of expansion, 
the proper motions of the three innermost knots (S0, S1, \& S2) 
would be expected to have either a strongly positive 
or negative right ascension term
if moving away from the pulsar or 
the remnant's center of expansion respectively. 
That is, they ought to move toward the southeast if 
expanding away from the pulsar (PA $\leqslant 180^{\circ}$) 
or toward the southwest if expanding away the 
Crab's expansion center (PA $\geqslant 180^{\circ}$). 

In contrast, all seven northern knots would be expected to show 
proper motions toward the northwest regardless of their 
origin point, thus making them a less powerful discrimination test 
without highly accurate positional measurements.
In addition, the innermost northern knot, N1, lying 
just $4''$ north of the pulsar, is a poorly defined feature and 
lies projected against a relatively bright filamentary 
background making precise positional measurements more difficult.

\subsection{Knot Proper Motions }

Table 2 lists the positions of the five southern knots 
(relative to the central star just northeast 
of the pulsar; Star 16 of \citealt{WM77}), our proper motion measurements, 
and with those cited by \citet{MacAlpine94}.
Within measurement uncertainites both data sets are in agreement. 
That is, although our estimated knot motions (except for Knot S1) 
are significantly smaller than those quoted by 
\citet{MacAlpine94}, they all lie within
the MacAlpine et al.'s relatively large error bar of 
$\pm0 \farcs 07$ yr$^{-1}$. 

However, it is puzzling why \citet{MacAlpine94} concluded
in favor of motions directed away from the pulsar with 
transverse velocities ``of order'' 900 km s$^{-1}$ 
based on their poorly constrained proper motion measurements.
Our east-west motions ($\delta$y values) of these five southern knots
are more than an order of magnitude below the \citet{MacAlpine94} detection limit.
But they are in line with simple radial motion away from
the center of expansion like that experienced by other remnant filamentary features.
A comparison of our knot proper motion estimates relative to the projected
displacements of the knots from the
center of expansion determine by \citet{Trimble68} is given in Table 3.
Here the average knot proper motions from the
$\delta$x and $\delta$y displacements
were calculated assuming these knots
experienced the same acceleration as the general
nebula leading to a derived explosion date
$\simeq$ A.D. 1130 \citep{Trimble68,WM77,Nugent98}.
As the table shows, our estimated knot proper motions 
are entirely consistent with the knots' projected 
radial displacements from the Crab's center of expansion.
[Note the somewhat better agreement using the \citet{Nugent98} expansion center estimate.]

Table 3 also lists the implied transverse velocities for the knots based on our
proper motion estimates, knot radial velocities reported by \citet{MacAlpine94}, 
and the resulting space velocities.
The five knots' space velocities range from 440 and 740 km s$^{-1}$. This 
is on the low end of the remnant's $700 - 2200$ km s$^{-1}$ 
filamentary expansion range \citep{DF85}.
This, in turn, places their location well inside the remnant's 
filamentary shell and thus relatively close to the 
X-ray and UV bright synchrotron nebula. 

Our proper motion estimates for all five knots 
indicate motions away from the remnant's expansion center and not the pulsar. 
Specifically, the position angles (PA) for Knots S0--S2 are all $>$ 180$^{\circ}$,
consistent with expected motions from either Trimble's or Nugent's 
estimated expansion centers. This is in contrast to the expected
PA values of around 152$^{\circ}$ if these knots were moving away from the pulsar. 
Our results are graphically shown in Figure 2 where we plot
the 100 yr proper motions of the five southern knots based on our measurements. 
As we found for the knots' total proper motion values, 
our measurements are in somewhat better agreement with the \citet{Nugent98} expansion center 
than that of \citet{Trimble68}. 

As a qualitative test of our results, 
we registered the 1994/95 images with the
1999 images and then blinked them.  We found the [Ar~III] knots' motions
to be visually consistent with radial motion away from the center of expansion
as opposed to away from the pulsar, supporting our quantitative
results.  We conclude the kinematics of these knots did not appear to be unusual
relative to other nebular features.

We believe our derived knot proper motions to be more reliable than
those of \citet{MacAlpine94} for several reasons. 
First, they used  ground-based [O~III] images 
with a much coarser pixel scale
($0 \farcs 46$) compared the {\it HST}  WFPC-2 images ($0 \farcs 1$) 
and with much lower image quality ($2''$ seeing
vs. WFPC-2's FWHM of $0 \farcs 045$). 
Second, their images covered just a 2.0 yr time span compared to our 5.66 yr (1994--1999) and
4.81 yr (1995--1999) time coverage. 

\subsection{Knot Morphology}

One criteria \cite{MacAlpine94} used to help identify [Ar~III] knots was a
semi-stellar and/or isolated appearance on their ground-based images. 
When examined using the higher resolution WFPC-2 images, however,
the knots described by \citet{MacAlpine94} show a broad range of 
morphologies and angular dimensions.
In Figure 3, we show enlargements of magnified WFPC-2 images 
for several of these [Ar~III] knots. 
While many clearly do have a somewhat stellar appearance with 
diameters of $\simeq$ $0\farcs4 - 0\farcs7$, others appear fairly 
diffuse with diameters of up to $1\farcs5$ (e.g., Knots N1, N6 \& S3). 

Figure 3 also shows enlarged magnifications of nine other 
nebular knots (labeled K1 -- K9) with sizes and morphologies similar to
the ``argoknots''. One of these, K7, was noted by 
\citet{MacAlpine94} as one of two other possible 
fainter and less distinct gas condensations.  
Similar knots are found throughout the central part 
of the remnant (see Fig. 1) and often lie near the north and south arcs of [Ar~III] knots.
The strong similarities in morphology of the [Ar III] knots to
these other nebular knots, and the placement of these other knots around the N-S knot arc, 
suggests that the [Ar III] knots are not unusual remnant features.

\section{Conclusions}

Our study of the proper motions of five [Ar~III] bright knots using
{\it HST} WFPC2 data from 1994 to 1999 show ordinary radial motions
away from the nebula's center of expansion with magnitudes consistent with
their projected radial displacement. The measurements indicate 
the knots are moving  
away from the remnant's expansion center and not the pulsar.
In addition, the [Ar~III] knots do not
appear to have unusual morphologies, and we have identified several other
similarly appearing knots in the remnant.
These results do not support the suggestion by \citet{MacAlpine94}
that these knots might be moving rapidly away from the Crab pulsar
due to a collimated pulsar-driven wind or that these knots are unusual
in appearance.

On the other hand, the spectral data of \citet{MacAlpine94} 
clearly show they possess strong [O~III] $\lambda\lambda$4959,5007 
with unusually strong [Ar~III] and [S~II] line emissions. 
MacAlpine et al. and \citet{Law95} noted a possible spatial 
association of the northern argo-knots with a filament at a higher 
radial velocity (+900 to +1300 km s$^{-1}$). The [Ar~III] bright knots
were seen mainly at lower-intensity breaks in the filament's emission.
They suggested that this might indicate that the material in the 
knots is derived from this filament possibly though some 
instability (Rayleigh-Taylor or Kelvin-Helmholtz).

Magnetic Rayleigh-Taylor instabilities at the interface
of this filament and the pulsar-generated synchrotron nebula
as discussed by \citet{Hester96} and \citet{Sankrit98} might well explain
the [Ar~III] knot morphologies.
The tips of the Rayleigh-Taylor, finger-like
filaments seen edge-on near the outer parts of the remnant
(e.g., Filaments F, G, \& H; \citealt{Hester96}) show
clumps of similar size and shape to the [Ar~III] knots.
The only difference here may be the viewing angle, 
where the [Ar~III] knots are seen face-on.

The observed difference in radial velocity between the coincident filament
and the chain of [Ar~III] knot would be consistent with
their more interior remnant positions. The estimated 440 -- 740 km s$^{-1}$ 
space velocities for these knots, at the extreme low end for the Crab's filaments,
is also consistent with this picture. This would
place them closer to the synchrotron nebula thus exposing them to
higher X-ray and UV photoionization fluxes. This, in turn, 
might be the underlying cause
for their strong [Ar~III] line emission via higher gas temperatures
as suspected by \citet{MacAlpine94}.

\acknowledgments

We thank V. Trimble for several helpful comments on the paper's
presentation. 

\clearpage

\begin{deluxetable}{lcccll}
\scriptsize
\tablecaption{WFPC-2 Images of the Crab Nebula}
\tablewidth{0pt}
\tablehead{
\colhead{Date} &  \colhead{Image} & \colhead{WFPC-2 } & \colhead{Emission } & \colhead{Region} &  \colhead{Exposures}  \\
\colhead{(U.T.)} & \colhead{ID}   & \colhead{Filter}  & \colhead{Line}      & \colhead{Imaged}  & \colhead{(s)}
}
\startdata
1994 Feb 23 & U24R0203T--204T &  F502N & [O III] & Center - NW  & 2000; 2000 \\
1994 Mar  9 & U24R0401T--402T &  F673N & [S II]  & Center - NW  & 2000; 2000 \\
1995 Jan 2 &  U2BX0401T--402T &  F673N & [S II] & Center - West & 2000; 2300   \\
1995 Jan 2 &  U2BX0403T--404T &  F673N & [S II] & Center - West & 2300; 2300   \\
1995 Jan 5 &  U2BX0301T--302T &  F502N & [O III] & Center - West  & 2000; 2300 \\
1995 Jan 5 &  U2BX0303T--304T &  F502N & [O III] & Center - West & 2300; 2300 \\
1999 Oct 24 & U5D10305R--306R &  F502N & [O III] & Center - SW  & 2600; 2600 \\
1999 Oct 24 & U5D10307R--308M &  F673N & [S II] & Center - SW  & 1300; 1300 \\
\enddata
\end{deluxetable}

\begin{deluxetable}{cclcccccc}
\tabletypesize{\small}
\footnotesize
\tablecaption{[Ar III] Knot Proper Motions}
\tablewidth{0pt}
\tablehead{ & \multicolumn{2}{c}{1995.0$^{a}$}   & \multicolumn{3}{c}{\underline{~ ~ ~ ~ ~ ~ ~ This paper ~ ~ ~ ~ ~ ~ ~}} & 
                 \multicolumn{3}{c}{\underline{~ ~ ~ ~ MacAlpine et al. ~ ~ ~ ~}}  \\
\colhead{Knot} & x & ~ ~ y  & \colhead{$\mu_{\rm x}$} & \colhead{$\mu_{\rm y}$} & \colhead{$\mu_{\rm tot}$} &
                 \colhead{$\mu_{\rm x}$} & \colhead{$\mu_{\rm y}$} & \colhead{$\mu_{\rm tot}$} \\
\colhead{ID~ } & ($''$) &~ ($''$)& \colhead{($''$ yr$^{-1}$)}& \colhead{($''$ yr$^{-1}$)} &\colhead{($''$ yr$^{-1}$)}  &
                 \colhead{($''$ yr$^{-1}$)} &  \colhead{($''$ yr$^{-1}$)} &\colhead{($''$ yr$^{-1}$)} } 
\startdata
S0  &  $3.5$ & $-16.7$  & $-0.007$  & $-0.011$ & 0.013$ \pm0.005$  &       \nodata  & \nodata   & \nodata \\
S1  &  $4.7$ & $-18.0$  & $-0.006$  & $-0.013$ & 0.014$ \pm0.004$  &     ~ 0.00     & ~ 0.00    &  $0.00 \pm0.07$ \\ 
S2  &  $6.2$ & $-21.5$  & $-0.006$  & $-0.017$ & 0.018$ \pm0.005$  &      +0.05     & $-0.05$   &  $0.07 \pm0.07$ \\
S3  & $11.9$ & $-27.3$  & $+0.004$  & $-0.021$ & 0.021$ \pm0.008$  &     ~ 0.00     & $-0.10$   &  $0.10 \pm0.07$ \\
S4  & $13.8$ & $-38.1$  & $+0.003$  & $-0.036$ & 0.036$ \pm0.007$  &     ~ 0.00     & $-0.10$   &  $0.10 \pm0.07$ \\
\enddata
\tablenotetext{a}{Positions relative to Star 16 of \citet{WM77}. }
\end{deluxetable}

\begin{deluxetable}{cccccccccc}
\tabletypesize{\small}
\footnotesize
\tablecaption{[Ar III] Knot Velocities and Directions }
\tablewidth{0pt}
\tablehead{
  & \colhead{$\Delta$x, $\Delta$y} &
    \multicolumn{5}{c}{\underline{~ ~ ~ ~ ~ ~ ~ ~ ~ ~ ~ ~ ~ ~ ~ This Paper~ ~ ~ ~ ~ ~ ~ ~ ~ ~ ~ ~ ~ ~}} &
    \multicolumn{3}{c}{\underline{Predicted $\mu$ Position Angles}} \\ 
\colhead{Knot} & \colhead{$\mu^{a}$} & \colhead{$\mu$}  &
\colhead{V$_{\rm t}^{\rm b}$} & \colhead{V$_{\rm r}^{c}$} &\colhead{V$_{\rm space}$} &
\colhead{PA} & \colhead{COE$^{\rm d}$} &  \colhead{COE$^{\rm e}$}   & \colhead{Pulsar} \\
\colhead{ID~ } &  \colhead{($''$ yr$^{-1}$)} & \colhead{($''$ yr$^{-1}$)} &
\colhead{(km/s)}  & \colhead{(km/s)}  & \colhead{(km/s)} &
\colhead{(deg)}  &  \colhead{(deg)}  & \colhead{(deg)}  & \colhead{(deg)}  }
\startdata
S0  & 0.011  & 0.013$ \pm0.005$  & 120  & \nodata  & \nodata & 215 $\pm16$ & 206 & 214  & 152  \\  
S1  & 0.012  & 0.014$ \pm0.004$  & 130  & 420      & 440     & 206 $\pm13$ & 197 & 205  & 151  \\ 
S2  & 0.015  & 0.018$ \pm0.005$  & 170  & 720      & 740     & 198 $\pm10$ & 187 & 194  & 152  \\
S3  & 0.022  & 0.021$ \pm0.008$  & 200  & 630      & 660     & 169 $\pm17$ & 167 & 173  & 147  \\
S4  & 0.035  & 0.036$ \pm0.007$  & 340  & 400      & 525     & 175 $\pm11$ & 168 & 172  & 153  \\
\enddata
\tablenotetext{a}{Proper motion values based on epoch 1995.0 knot displacements ($\Delta$x, $\Delta$y) 
                  from Trimble's center of expansion and assuming an expansion date of A.D. 1130.}
\tablenotetext{b}{Average transverse velocity estimates based on our proper motion 
  measurements and a 2 kpc distance.}
\tablenotetext{c}{Radial velocities from \citet{MacAlpine94}. }
\tablenotetext{d}{Measured from Trimble's (1968) center of expansion (COE).}
\tablenotetext{e}{Measured from Nugent's (1998) center of expansion (COE).}
\end{deluxetable}

% This is the last table for this paper (as well as the first), so we
% should follow it with a \clearpage.  In order to force all the floating
% tables out of their buffers and onto vertical page lists, we must use
% \clearpage rather than \newpage. 

% This is the last table for this paper (as well as the first), so we
% should follow it with a \clearpage.  In order to force all the floating
% tables out of their buffers and onto vertical page lists, we must use
% \clearpage rather than \newpage. 

\clearpage

% Now comes the reference list.  In this document, we used \cite to call
% out citations, so we must use \bibitem in the reference list, which
% means we use the LaTeX thebibliography environment.  Please note that
% \begin{thebibliography} is followed by a null argument.  If you forget
% this, mayhem ensues, and LaTeX will say "Perhaps a missing item?" when
% you run it.  Do not call us, do not send mail when this happens.  Put
% the silly {} after the \begin{thebibliography}.
%
% Each reference has a \bibitem command to define the citation format
% to be placed in the text (in []) and the symbolic tag used for 
% cross referencing (in {}).
%
% See sample1.tex, or the AASTeX guide, for an alternative to the \cite-
% \bibitem command.

\clearpage
\newpage

\clearpage
\begin{figure*}
\begin{center}
\includegraphics[scale=0.70,clip=true,trim=0in 0.0in 0in 0in]{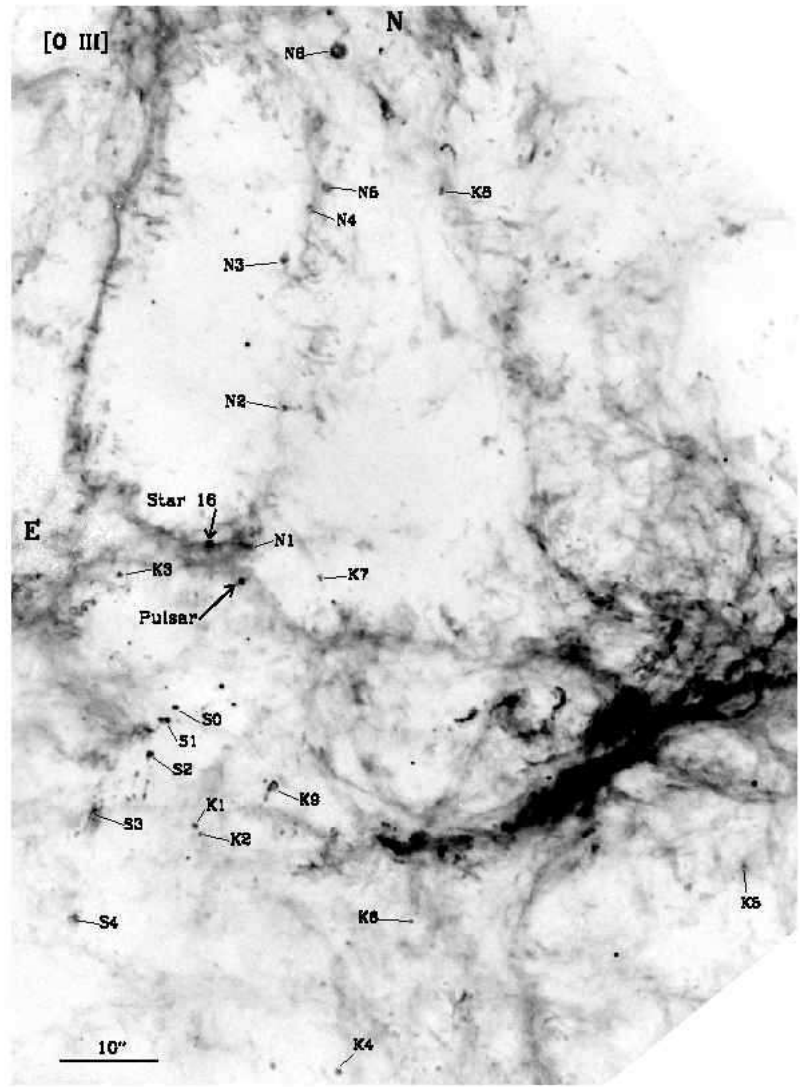}
\caption{1995 {\it HST} [O~III] $\lambda$5007 WFPC2 image of the west-central region 
of the Crab Nebula showing the line of [Ar III] emitting knots (N1-N6 and S1-S4) 
described by \citet{MacAlpine94}. The field of view shown is approximately $1.3' \times 1.8'$.
Other nebular knots having similar ``semi-stellar'' 
[O~III] morphologies are shown in the image and labeled K1 -- K9. }
\label{fig1}
\end{center}
\end{figure*}

\clearpage

\begin{figure*}
\begin{center}
\includegraphics[scale=1.00,clip=true,trim=1.0in 1.5in 0in 2in]{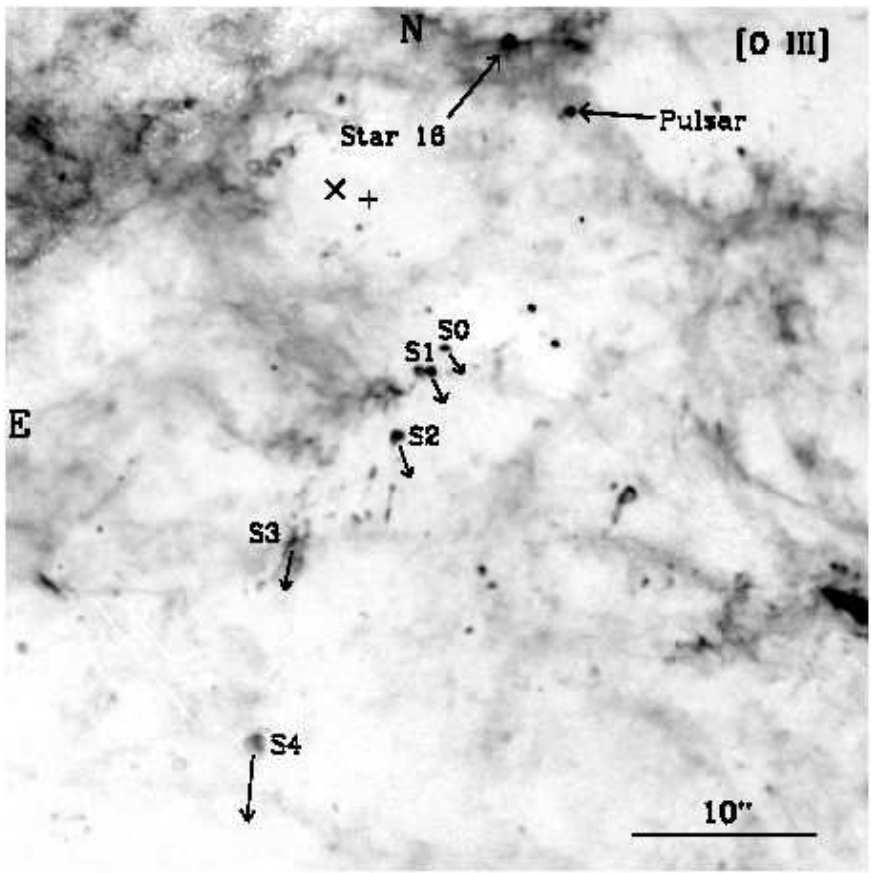}
\caption{1995 {\it HST} [O~III] $\lambda$5007 WFPC2 image of the 
center of the Crab Nebula showing the directions
of motion of the five southern knots.  
The length of the arrows corresponds to the distance
that would be traveled in 100 years at the present velocity.  
The cross marks the position of the \citet{Trimble68} center of expansion, 
the x the position of the \citet{Nugent98} center of expansion.} 
\label{fig2}
\end{center}
\end{figure*}

\clearpage

\begin{figure*}
\begin{center}
\includegraphics[scale=0.78,clip=true,trim=0in 0.5in 0in 0in]{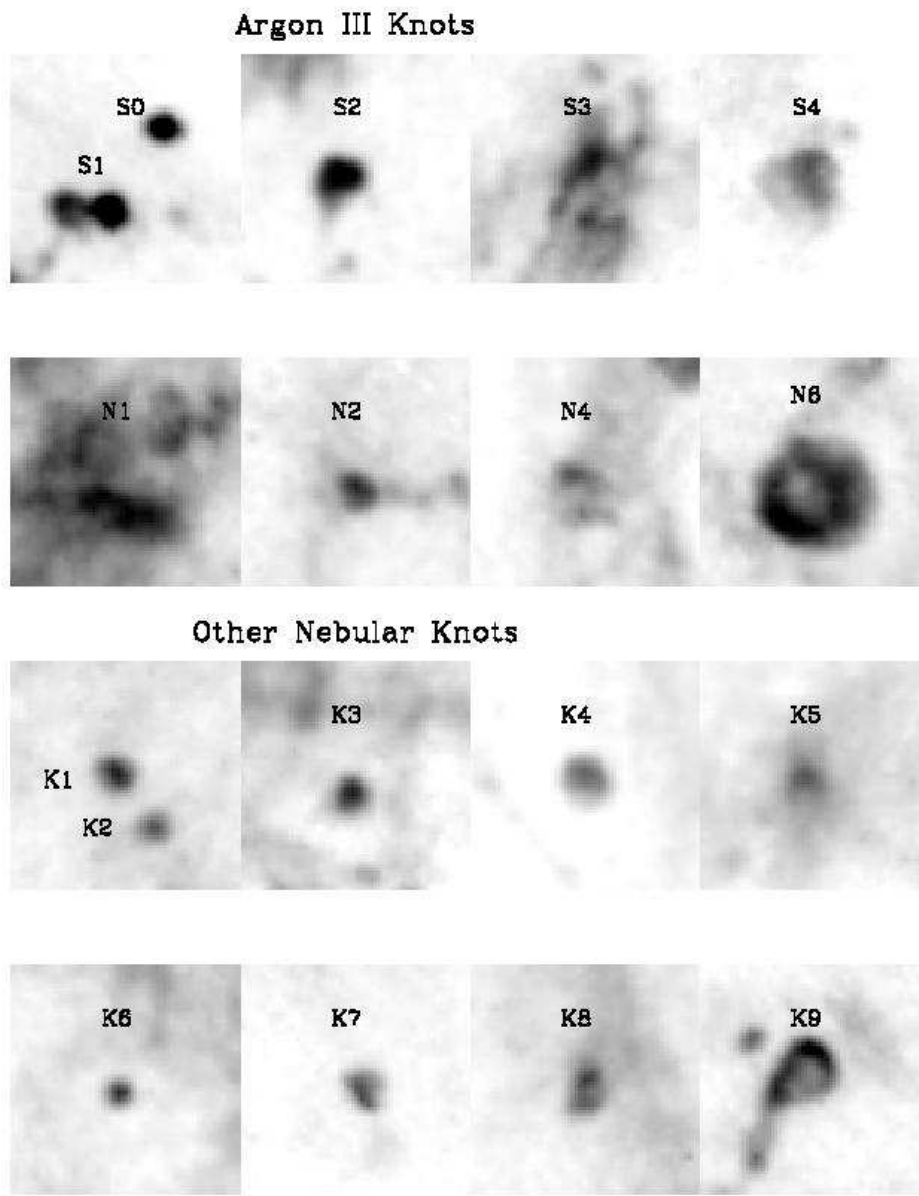}
\caption{1995 {\it HST} [O~III] $\lambda$5007 WFPC-2 
magnified ($0\farcs05$/pixel) images of bright [Ar III]
emitting knots and of other nebular knots.  
Each box is 5$''$ square.}
\label{fig3}
\end{center}
\end{figure*}


\begin{thebibliography}{}
\bibitem[Blair et al.(1997)]{Blair97}Blair, W. P., Davidson, K., Fesen, R. A., 
        Uomoto, A., MacAlpine, G. M., \& Henry, R. B. C. 1997, \apjs, 109, 473
\bibitem[Davidson \& Fesen(1985)]{DF85}Davidson, K., \& Fesen, R. A. 1985, \araa, 23, 119
\bibitem[Fesen et al.(1992)]{Fesen92} Fesen, R. A., Martin, C. L., \& Shull, J. M.
         1992, \apj, 399, 599
\bibitem[Hester et al.(1996)]{Hester96} Hester, J.\ J.\ et al.\ 
           1996, \apj, 456, 225
\bibitem[Lawrence, \& MacAlpine(1994)]{LM94} Lawrence, S. S., \& MacAlpine, G. M.
         1994, \baas, 26, 951 
\bibitem[Lawrence et al.(1995)]{Law95}Lawrence, S. S., MacAlpine, G. M., Uomoto, A.,
           Woodgate, B. E., Brown, L. W., Oliversen, R. J., Lowenthal, J. D., \&
            Liu, C. 1995, \aj, 109, 2635
\bibitem[Kafatos \& Henry(1985)]{KH85}Kafatos, M. C., \& Henry, R. B. C. 1985,
        ``The Crab Nebula and Related Supernova Remnants - An Overview'',
         (Cambridge, Cambridge Univ. Press)
\bibitem[MacAlpine et al.(1989)]{MMMU89} MacAlpine, G. M., McGaugh, S. S.,
        Mazzarella, J. M., \& Uomoto, A. 1989, \apj, 342, 364
\bibitem[MacAlpine, Lawrence, \& Brown(1993)]{MacAlpine93} MacAlpine, G. M., Lawrence, S. S.,
          \& Brown, B. A. 1993, \baas, 25, 785
\bibitem[MacAlpine et al.(1992)]{MacAlpine92}MacAlpine, G. M., Lawrence, S. S., 
        Uomoto, A., Woodgate, B. E., Brown, L. W., Oliversen, R. J.,
        Lowenthal, J. D., \& Liu, C. 1992, \baas, 24, 791
\bibitem[MacAlpine et al.(1994)]{MacAlpine94}MacAlpine, G. M., Lawrence, S. S., Brown, B. A.,
        Uomoto, A., Woodgate, B. E., Brown, L. W., Oliversen, R. J.,
        Lowenthal, J. D., \& Liu, C. 1994, \apj, 432, L131
\bibitem[Michel et al.(1991)]{Michel91}Michel, F. C., Scowen, P. A.,
         Dufour, R. J., \& Hester, J. J. 1991, \apj, 368, 463
\bibitem[Nugent(1998)]{Nugent98} Nugent, R. L. 1998, \pasp, 110, 831
\bibitem[Trimble(1968)]{Trimble68}Trimble, V. 1968, \aj, 73, 535
%\bibitem[Trimble(1973)]{Trimble73}Trimble, V. 1973, \pasp, 85, 579
\bibitem[Trimble(1985)]{Trimble85}Trimble, V. 1985, in ``The Crab Nebula and Related Supernova
         Remnants - An Overview'', (Cambridge, Cambridge Univ. Press), p 257
\bibitem[Sankrit et al.(1998)]{Sankrit98} Sankrit, R.\ et al.\ 
          1998, \apj, 504, 344
\bibitem[Uomoto \& MacAlpine(1987)]{UM87}  Uomoto, A., \& MacAlpine, G. M. 1987, \apj, 160 L27
\bibitem[van den Bergh(1970)]{vdb70} van den Bergh, S. 1970, \apj, 160, L27
\bibitem[Wyckoff \& Murray(1977)]{WM77}Wyckoff, S., \& Murray, C. A. 1977, \mnras, 180, 717
\end{thebibliography}
\end{document}